\documentclass{pnastwo}

\usepackage{pnastwof}
\usepackage{amssymb,amsfonts,amsmath}
\usepackage{sidecap}
\usepackage{color}
 
\usepackage{graphicx}
\usepackage{amssymb}
\usepackage{epstopdf}
\usepackage[extension=xxx]{hyperref}
\newcommand{\unit}[1]{\ensuremath{\, \mathrm{#1}}}

\pdfoutput=1

\def\be{\begin{equation}}   \def\ee{\end{equation}}
    \def\fig#1{{Fig.\ref{#1}}}
\def\kT{k_{{}_{\rm B}}T}

  \def\koff0{{k_{\rm off,0}}}

\begin{document}

\title{Quantitative analysis of intra-Golgi transport reveals inter-cisternal exchange for all cargo}
\author{Serge Dmitrieff \affil{1}{Laboratoire Gulliver, CNRS-ESPCI, UMR 7083, 10 rue Vauquelin, 75231 Paris Cedex 05, France, pierre.sens@espci.fr}, Madan Rao \affil{2}{Raman Research Institute, C.V. Raman Avenue, Bangalore 560 080, India} \affil{3}{National Centre for Biological Sciences (TIFR), Bellary Road, Bangalore 560 065, India} and Pierre Sens\affil{1}{ } }

\contributor{Submitted to Proceedings of the National Academy of Sciences
of the United States of America}

\maketitle

\begin{article}

\noindent
BIOLOGICAL SCIENCES : Cell Biology
\\

\begin{abstract}
The mechanisms controlling the transport of proteins across the  Golgi stack of mammalian and plant cells is the subject of intense debate, with two models,  cisternal progression and inter-cisternal exchange, emerging as major contenders.
A variety of transport experiments have claimed support for each of these models. We reevaluate these experiments using a single quantitative coarse-grained framework of intra-Golgi transport that accounts for both transport
models and their  many variants.
Our analysis makes a definitive case for the existence of inter-cisternal exchange both for small membrane proteins (VSVG) and large protein complexes (procollagen) -- this  implies that membrane structures larger than the typical protein-coated vesicles must be involved in transport. Notwithstanding, we find that current observations on protein transport cannot rule out cisternal progression as contributing significantly to the transport process.
To discriminate between the different  models of intra-Golgi transport, we suggest experiments and an analysis based on our extended theoretical framework that compare the dynamics of transiting and resident proteins.
\end{abstract}


\abbreviations{EM: Electron Microscopy, ER: Endoplasmic Reticulum, FRAP: Fluorescent Recovery After Photobleaching, PMC: Pleiomorphic membrane carrier}

\begin{figure}[t]   
\centering
\includegraphics[width=1.0\linewidth]{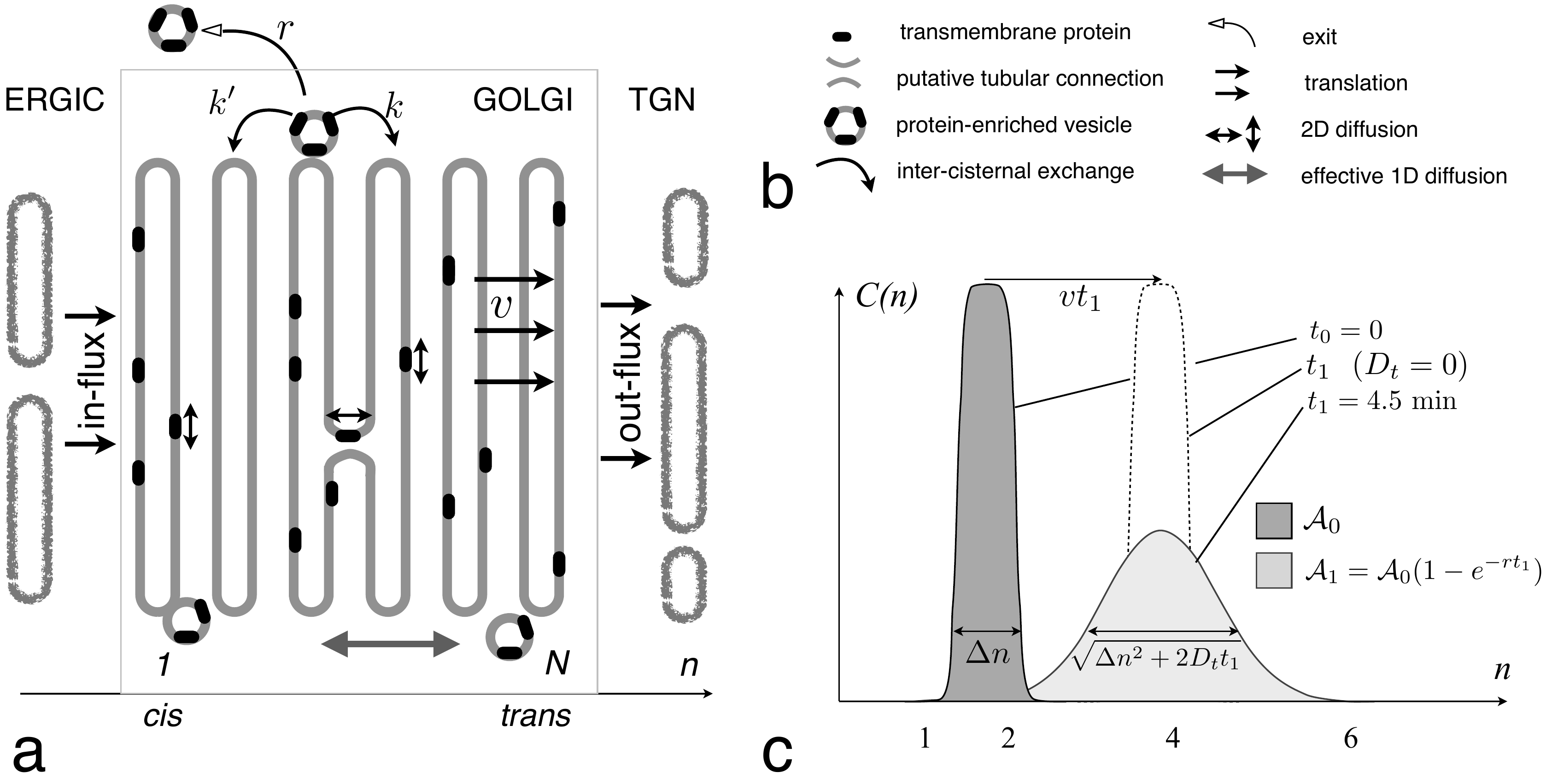}
\caption{\label{schemagolgi} \small {Sketch of the Golgi apparatus as a polarized stack of connected cisternae exchanging material. \bf (a)} Proteins synthesized in the endoplasmic reticulum (ER) go through the so-called ER-Golgi intermediate compartment (ERGIC) before entering the Golgi through its  {\em cis} face. After biochemical maturation and sorting, they exit the Golgi through the {\em trans} face to join the trans-Golgi-network (TGN) and are dispatched to particular cellular locations. {\bf (b)} Relevant transport processes, including cisternal progression (translation), diffusion through connecting membrane tubules, bidirectional vesicular transport, and exit. {\bf (c)} Spatio-temporal evolution of an initially narrow protein distribution (as produced by a pulse of secretion from the ER); pure convection produces a uniform translation of the peak (dashed line), diffusion broadens the peak, and exit exponentially decreases the protein content.}
\end{figure}

\dropcap{T}he Golgi apparatus, a complex cellular organelle responsible for lipid and protein maturation and sorting, has attracted a lot of attention, with many conflicting viewpoints regarding its  mechanisms of transport. The Golgi of plant and animal cells  consists of a stack of  $5$ to $20$ cisternae\cite{polishchuk:2004}, possibly interconnected by membrane tubules \cite{marsh2004direct}, which exchange material by vesicle budding and fusion \cite{orci1989dissection,malhotra1988role} (see \fig{schemagolgi}).  Each cisterna has a distinct chemical identity, allowing progressive protein maturation from the {\em cis} to the {\em trans}  face \cite{wilson:2010}. 

There is a long standing argument about the way proteins are transported through the Golgi, an issue intimately tied to the structure and dynamics of the organelle itself. The Golgi could be a rather static structure, in which  cisternae keep constant positions and identities, and exchange proteins by vesicular transport. Alternatively, cisternae could progress from the {\em cis} end to the {\em trans} end without exchanging their cargo \cite{marsh:2002}. Biochemical maturation of individual cisterna is known to occur in yeast (Saccharomyces cerevisiae) Golgi, which is not stacked but made of dispersed cisternae \cite{Matsuura:2006,losev:2006}. The cisternal progression model posits that this maturation translates into a physical progression of the cisternae (and their content) along the stack. It is supported by the observation that large molecules such as procollagen aggregates, presumably unable to enter conventional transport vesicles, nonetheless progress through the stack, suggesting that cisternae are created at the {\em cis} face and destroyed at the {\em trans} face \cite{bonfanti:1998}. This picture was recently challenged  \cite{patterson:2008} by the observation that  proteins do not exit the Golgi linearly with time (as a model purely based on cisternal progression would predict) but exponentially, as can be explained by  inter-cisternal exchange. These are however two extreme models, and 
 cisternal progression and inter-cisternal exchange could act concomitantly. This is clear even in the cisternal progression model, which
  requires that resident Golgi enzymes (which are found in particular location in the Golgi stack) undergo specific retrograde (vesicular) transport.

The relevance of each transport phenomenon for a given protein species can  be properly evaluated only by confronting  experimental observations with an unbiased quantitative model based on general physical principles. Existing models do not adopt this approach. They are often tailored to support \cite{glick:1997} or disprove \cite{patterson:2008} the cisternal progression model, and their comparison with quantitative data involves a large number of fitting parameters \cite{patterson:2008}. 

Recent advances in super-resolution microscopy lends hope that the spatio-temporal distribution of proteins inside the Golgi may soon be resolved. This  calls for a rigorous description of intra-Golgi transport based on the general formalism of transport phenomena \cite{vankampen:2007}. We describe such a framework, where transport is characterized by generic coarse-grained transport rates. These parameters can be related to microscopic processes using specific models, but the framework itself is largely  model-independent. We report here that all available quantitative data on a variety of cargo, including large procollagen aggregates, can be reproduced by a 
combination of (i) global protein translation from the {\em cis} to the {\em trans} Golgi, (ii) diffusive-like protein exchange between cisternae, and (iii) protein exit throughout the stack. As shown below, the diffusive component  implies that inter-cisternal exchange is not restricted to small protein-coated vesicles, and involves large transport carriers. We rigorously establish that transport data based on 
tagging a single molecular species can be argued to be consistent with many different models of transport and therefore cannot provide an unequivocal picture of intra-Golgi transport. To reach this goal, we propose experimental strategies based on dynamical correlations between transiting and resident Golgi proteins. A useful virtue of our formalism is that it can include the influence of the local biochemical and physical environment within the different cisternae as an energy landscape through which proteins diffuse, and thus permits a description of transiting proteins and resident Golgi enzymes within the same mathematical framework. 

\section{Model}\label{model}

\subsection{Transport equations  for inter-cisternal exchange}\label{discrete}

Treating the Golgi stack as composed of distinct cisternae, we analyze protein transport along its axis of polarity (the {\em cis}-{\em trans} axis), for which the cisterna number $n$, varying between $1$ (the {\em cis}-most) to $N$ (the {\em trans}-most) plays the role of a discrete spatial coordinate. The distribution of a chemical species $A$ within the Golgi may be characterized by its concentration $A_n(t)$ in cisterna $n$ at time $t$. Inter-cisternal exchange is restricted to ``jumps'' between adjacent cisternae (with rates $k_n$ for $n\rightarrow n+1$ and $k_n'$ for $n\rightarrow n-1$, see \fig{schemagolgi}). We emphasize that the rates $k_n,\ k'_n$ and $r_n$ characterizing the coarse-grained dynamics may be used {\it regardless} of the microscopic details of the exchange process. For vesicular transport, they are the product of the rates of fission, translocation and fusion of vesicles carrying $A$, and  include the waiting time of $A$ within a cisterna. They are not restricted to processes involving protein-coated vesicles, and may include transport through connecting membrane tubules and contributions from any fragment that detaches from one cisterna and fuse with a neighboring cisterna. In general, these rates do not obey detailed balance, and can in principle depend on the local concentration $A_n$. 
A ``master equation''  \cite{vankampen:2007} can be written for the 
concentration $A_n(t)$:
\begin{eqnarray}
\partial_t A_n(t)&=\underbrace{k_{n-1}A_{n-1}-k'_nA_n}_{\text{net flux:\ }n-1\rightarrow n}-(\underbrace{k_nA_n- k'_{n+1}A_{n+1}}_{\text{net flux:\ }n\rightarrow n+1})
\label{kindiscrete}
\end{eqnarray}
A straightforward  generalization of the model could include transport between distant cisternae. This however does not bring new insight, nor does it improve the comparison with available experimental data on transiting proteins. 

We will rewrite Eq.\eqref{kindiscrete} in a continuous formalism,  since this allows for a better description of cisternal progression. The coordinate $n$ (the cisterna number) can be written as a continuous variable, and spatial variations are then written as a derivative: $\partial_n A_n=(A_{n+1}-A_{n-1})/2$, with distances normalized by the inter-cisternal distance (the connection between the discrete and continuous models is described in detail in the Supplementary Information - S.I). If the different exchange rates do not depend too drastically on position ($\partial_nk_n\ll k_n$), Eq.\eqref{kindiscrete} can be transformed into the so-called Fokker-Planck equation \cite{vankampen:2007} (we show below that the continuous approximation is appropriate for the existing experimental data). In this continuous description, inter-cisternal exchange amounts to an effective translation with velocity $v_t$, combined with an effective diffusion with a diffusion constant $D_t$:
\begin{eqnarray}
\frac{\partial A_n}{\partial t} =\frac{\partial}{\partial n}\left( D_t\frac{\partial A_n}{\partial n} -v_t  A_n\right)\nonumber \\
{\rm with}\quad D_t = \frac{k_{n}+k'_{n+1}}{2} \quad {\rm and} \quad v_t=k_{n}-k'_{n+1}
\label{kineq_ex}
\end{eqnarray}
This illustrates that inter-cisternal exchange always yields an effective diffusion coefficient, even if all transport steps are anterograde ($k_n>0 \ , \ k_n'=0$), as we discuss below. 

\subsection{Including cisternal progression and external fluxes}\label{continuous}

Proteins may  be transported toward the Golgi {\em trans} face by ``cisternal progression'', defined as the process by which the entire content of a cisterna moves from position $n$ to position $n+1$ in the stack over a time $\Delta t$. The ``progression velocity'' is thus defined as $v_p\equiv1/\Delta t$, and is the same for all cisternae. Furthermore, the species $A$ may in principle be imported to or exported from any cisterna along the stack.  These processes, which include direct recycling to the ER, may be expressed as an external flux $J^n$ composed of an influx $J_{\rm in}^n$ to cisterna $n$ and a rate of exit $r_n$ from cisterna $n$. Eq.\eqref{kineq_ex} becomes:
\begin{eqnarray}
\frac{\partial A_n}{\partial t}& =&\frac{\partial}{\partial n}\left( \overbrace{D_t\frac{\partial A_n}{\partial n}}^{\text{diffusion}} -\overbrace{\vphantom{D_t\frac{\partial A_n}{\partial n}}(v_p+v_t)  A_n}^{\text{net translation}}\right) + \overbrace{\vphantom{D_t\frac{\partial A_n}{\partial n}}J^n}^{\text{external flux}}\label{kineq}\\
&&{\rm with}\quad J^n=J_{\rm in}^n-r_nA_n \nonumber 
\end{eqnarray}
The influx $J_{\rm in}^n$ could come from outside the Golgi, or could originate from distant cisternae (in which case it depends on the concentration in these cisternae).
Since it is not expected to contribute significantly to the dynamics of transiting proteins coming from the ER, it is ignored for now but is revisited in the Discussion section when we comment on
 the distribution of resident Golgi enzymes. Fluxes entering at the {\em cis} face and exiting from the {\em trans} face of the stack are included in the model as boundary fluxes (see below).

Eq.\eqref{kineq} illustrates  three fundamental mechanisms governing the temporal evolution of a protein distribution within  the Golgi:  
 {\em i)} protein exchange between neighboring cisternae introduces an effective diffusion of the concentration along the Golgi stack, characterized by a diffusion coefficient $D_t$, {\em ii)} directed protein transport from the {\em cis} to the {\em trans} Golgi leads to protein translation at a velocity $v=v_t+v_p$, this accounts both for cisternal progression (at velocity $v_p$) and for a bias for anterograde ($v_t>0$) or retrograde ($v_t<0$) inter-cisternal exchange, and {\em iii)} proteins may in principle exit from any Golgi cisterna to join other organelles (the ER or lysosomes) at a rate $r_n$, which may be zero. Note that since the spatial coordinate is a dimensionless number, all three parameters have units of rates ($s^{-1}$).

Because it does not depend on the microscopic processes responsible for transport, Eq.\eqref{kineq} constitutes the most rigorous quantification of an arbitrary transport process, and should be used as a first approach to characterize Golgi transport.
The impact of the three main parameters on the distribution of proteins throughout the Golgi is best seen when analyzing the propagation of an initially localised protein distribution (pulse-chase experiments, \fig{schemagolgi}). The translation velocity displaces the concentration peak (linearly in time if $v$ is constant), diffusion broadens the peak (its width  increases as the square root of time if $D_t$ is constant) and protein exit decreases the total protein concentration (exponentially with time if $r$ is constant). The various rates could vary for different proteins, possibly transported by different mechanisms, and should in particular be very different for transiting proteins and resident Golgi enzymes. 

Cisternal progression only affects the translation velocity in Eq.\eqref{kineq}, while anterograde inter-cisternal exchange affects both the velocity and the diffusion coefficient. Our formalism thus readily shows a fundamental qualitative difference between the two contending models.  Within the cisternal progression model, the movement of transiting proteins may occur in the absence of inter-cisternal exchange,  thus $v_p>0$, $k_n=k'_n=0$, which amounts to a perfect translation, without broadening, of a peak of concentration , i.e. $D_t=0$. Inter-cisternal exchange, on the other hand, necessarily involves some broadening, with an apparent diffusion coefficient directly related to the translation velocity ($D_t=v_t/2$ in the absence of retrograde transport, i.e., when $k_n'= 0$, and $D_t>v_t/2$ if $k_n'\ne 0$). This immediately leads to a powerful conclusion: if the analysis of the pulse chase data using Eq.\eqref{kineq} suggests that $v>2 D_t$, then we can unambiguously conclude that the data is incompatible with a transport based purely on inter-cisternal exchange and must allow for {\it some} cisternal progression.
 This illustrates how  a  quantitative analysis based on generic transport equations may shed light on the nature of intra-Golgi transport, without requiring the knowledge of  microscopic details of individual transport steps. Microscopic details do however control the values of the different rates, and whether they vary along the Golgi stack or are influenced by the presence of other proteins.  It is known for instance that cargoes can influence their own transport, in particular by interacting with the COP machinery responsible for vesicle formation \cite{liu:2005,forster:2006}.  We  show in the next section that all available data for transiting proteins are well fitted by assuming constant exchange rates.  
 A significant feature of this framework is that it is easily generalizable. For instance, 
 spatial variation of the transport rates can be easily incorporated within our framework. We show in the discussion section and in the S.I. that information on the biochemical and physical environment of individual cisternae can be prescribed using an energy landscape formalism.

\subsection{Boundary Fluxes}\label{BC}
Eq.\eqref{kineq} must be supplemented by boundary conditions  at the  {\em cis} ($n=1$) and {\em trans}  ($n=N$) faces of the stack. At the {\em cis} face, the influx of material $J_{\rm in}^1$ from the ER 
is  taken as a  parameter (possibly varying with time), imposed by the experimental procedure ({\em e.g.,} in pulse-chase or incoming wave protocols, see below). The rate of protein exit at the {\em cis} face is taken as a fitting parameter  $k^-$($=k'_{n=1}$). The out-flux of material at the {\em trans} face $J_{\rm out}^N$ includes contributions both from vesicles secreted at the {\em trans} Golgi and from the maturation of the {\em trans} cisterna: $J_{\rm out}^N=(v_p+k_N)A_N$. As can be seen from Eq.\eqref{kineq} these two contributions may not be easily distinguished, as the net flux throughout the Golgi involves the net velocity $v=v_p+v_t$. We thus write the exit flux $J_{\rm out}^N=(v+k^+)A_N$, where $k^+=k_N-v_t$ is the fitting parameter of {\em trans} Golgi exit. 
In addition to the transport parameters ($v$ and $D_t$) and the exit rate $r$, there are thus two additional boundary parameters $k^-$ and $k^+$ in the model. Boundary conditions do affect the spatio-temporal distribution of proteins inside the Golgi, but we show below that the (bulk) parameters $D_t$ and $v$, which control the actual transport through the Golgi, can nevertheless be determined with reasonable accuracy.

\section{Results}\label{rrr}

 \begin{figure*}[b]
 \caption{\label{fig-exp} \small 
Quantitative analysis of data from different experimental protocols using a numerical solution of Eq.\eqref{kineq}. }
 \begin{tabular}{cc}
 \begin{minipage}{0.65\linewidth}
\centerline{\includegraphics[width=11.5cm]{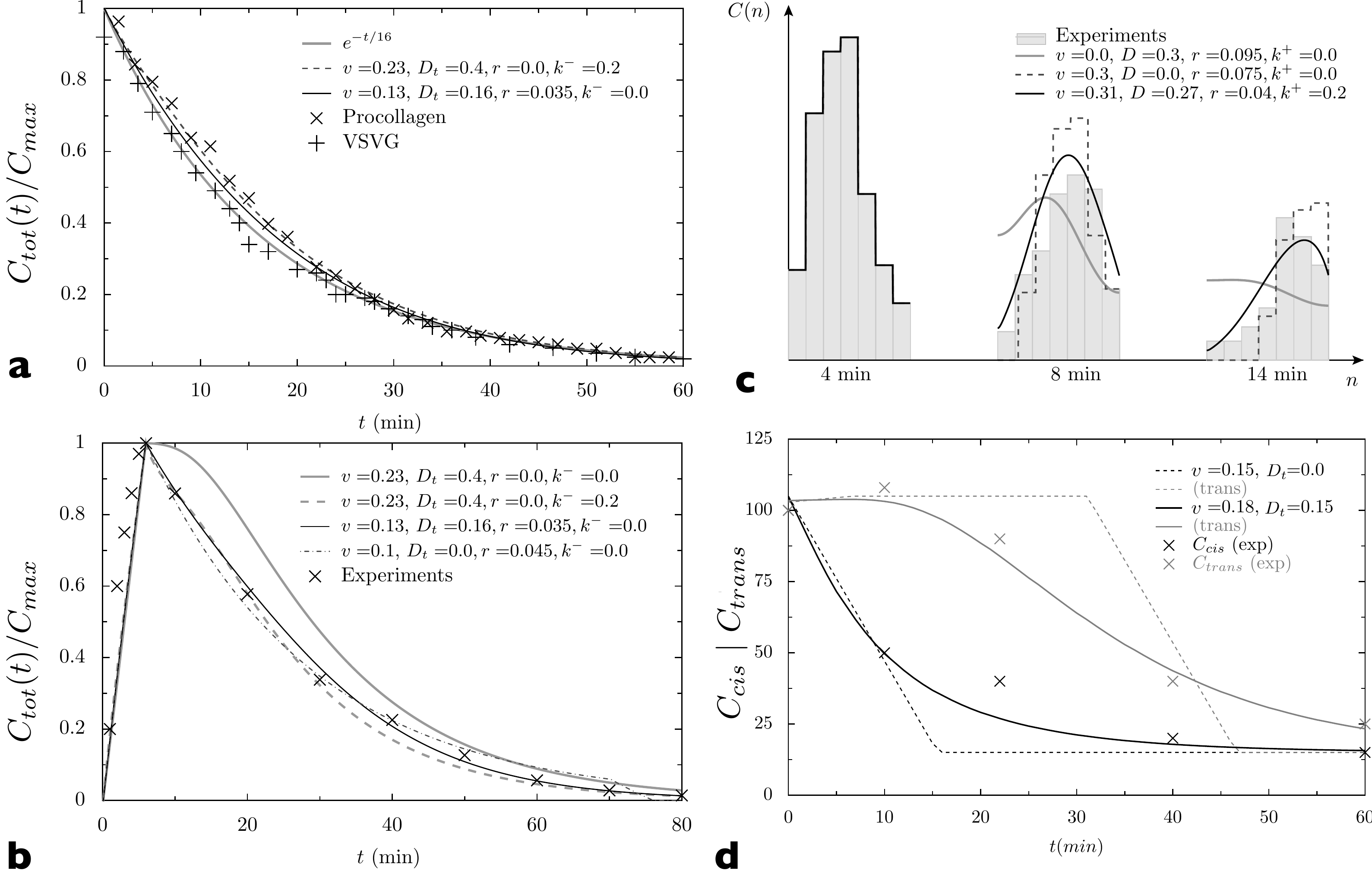}}
\end{minipage}
& 
\begin{minipage}{0.35 \linewidth} \small \vspace{0.3pt}
{\bf (a, b)} Optical microscopy assays. A whole Golgi FRAP experiment probing the exit of tagged proteins from the Golgi following {\bf (a)} a steady influx, abruptly stopped at $t=0$, of a small transmembrane protein (glycoprotein of vesicular stomatitis virus - VSVG) and a large soluble protein aggregate (procollagen), and {\bf (b)} a short influx, starting at $t=0$ and stopping at $t=5\unit{min}$, of  VSVG  \cite{patterson:2008}. $k^+$ was set to zero in the fits because it does not influence the early relaxation.
{\bf (c, d)}  Electron microscopy assays. {\bf (c)} A pulse chase experiment for VSVG, \cite{trucco:2004}. Setting either convection (grey curve) or diffusion  (dashed curve) to zero cannot reproduce the data.  Fits are constrained so that the total protein concentration matches the data at $t=14\unit{min}$. $k^-$ was set to zero because it has the same effect as $r$. {\bf (d)} Evolution of the concentration of procollagen aggregates in the {\em cis} (black) and {\em trans} (grey) face of the Golgi upon sudden blockage of ER secretion (exiting wave experiment) \cite{bonfanti:1998}. Data are in percentage of the concentration in normal conditions (steady ER secretion), and are not sensitive to exit rate.
 More information on the fitting procedure and experimental uncertainty is given in the S.I.  \end{minipage}
  \end{tabular}
\end{figure*}

\subsection{Confrontation with experimental data on transiting proteins}\label{exp}

Our theoretical framework was used to analyze different experimental observations, collectively illustrated in \fig{fig-exp}.
Fluorescence Recovery After Photobleaching experiments (FRAP) performed on the whole Golgi gives access to the total concentration of tagged proteins inside the Golgi. An exponential  recovery dynamics is reported in \cite{patterson:2008},  both for small membrane proteins (VSVG) and for large cytosolic protein complexes (procollagen). This was used as an argument against pure cisternal progression, for which a linear recovery dynamics is expected. 

Our analysis shows  (\fig{fig-exp}a) that the recovery profile is rather insensitive to the mode of intra-Golgi transport, and in particular to the effective diffusion coefficient $D_t$, the only parameter that {\it solely} depends on inter-cisternal exchange.  We fit the data with a single exponential decay of characteristic time $16$ min, which could be accounted for by any one of the following; protein exit throughout the Golgi (parameter $r$), early exit from the {\em cis} face (parameter $k^-$), late exit via the {\em trans} face (parameter $k^+$), or any combination of the three. The dynamics of small inert soluble cargo molecule reported in \cite{patterson:2008} follows a similar, although slightly faster, exponential recovery, with similar conclusions regarding its means of transport. When fluorescent VSVG proteins were only allowed to enter the Golgi for a short time, the exponential recovery started immediately after the cessation of the fluorescence in-flux (\fig{fig-exp}b). This shows that proteins do not need to reach the {\em trans}  face to exit the Golgi, since recovery would otherwise show a delay (grey  curve in \fig{fig-exp}b), and suggest that proteins can exit at the {\em cis} face (parameter $k^-$) or throughout the stack (parameter $r$).  
Although these experiments give important information concerning the rate at which proteins are exported from the Golgi, such average measures of Golgi dynamics  do not yield  any clear-cut conclusion on the dominant means of  transport across the Golgi stack. For instance, the exponential fluorescence decay of both FRAP experiments is consistent with a transport solely based on cisternal progression ($D_t=0$), provided proteins are allowed to exit throughout the Golgi  at a sufficient rate ($r \gg v/N$). A quantitative assessment of intra-Golgi transport, which is tantamount to obtaining numerical values for $v$ and $D_t$, requires the knowledge of the protein distribution inside the entire organelle.
 
 Following the transport of a pulse of  protein (pulse-chase protocol - \fig{schemagolgi}c), or  the evolution of the protein distribution across the Golgi after ER secretion has been suddenly blocked (exiting wave protocol), could in principle yield independent measurements of the various parameters. Our analysis of pulse-chase data on small membrane proteins (VSVG, \fig{fig-exp}c) \cite{trucco:2004} clearly shows a combination of translation ($v\ne 0$), broadening ($D_t\ne 0$) and decay (at least one non-vanishing parameter among $\{r,k^-,k^+\}$)  of the peaked concentration distribution. The best fit (black curve in \fig{fig-exp}c) suggests that all transport rates have similar values ($v\sim D_t\sim k^+\simeq 0.2-0.3$ min$^{-1}$. The value of the velocity corresponds to a transit time across the Golgi of order $t_{\rm transit}=N/v\simeq15$ min (where $N\simeq 6$ is the number of cisternae). More importantly, the high value of the diffusion coefficient indicates that VSVG is exchanged between cisternae during its transport through the Golgi. 
 
 For large cytosolic procollagen aggregates, the exiting wave protocol reported in \cite{bonfanti:1998} shows that concentration differences between the {\em cis} and {\em trans} Golgi relax rather smoothly after secretion is stopped, unlike what would be expected within a pure cisternal progression model (solid lines compared to dashed lines in \fig{fig-exp}d). Our analysis of the (rather scarce) data suggests that,  just as VSVG, procollagen undergoes inter-cisternal exchange with a fairly large diffusion coefficient, $D_t\simeq v$. This large value of $D_t$ is rather surprising for such large protein complexes and of fundamental significance.
 
Experimental limitations, such as variability within and between cells or the finite amount of time needed to set up  transport block, could be argued to smoothen concentration gradients in a way similar to inter-cisternal exchange. We show in the S.I. that given the experimental error (below $10\%$ for data of \fig{fig-exp}d, \cite{bonfanti:1998}), a finite diffusion coefficient must be invoked to explain the procollagen exiting wave data  provided ER export ceases within $10\unit{min}$ of the initiation of the block. For a $5\unit{min}$ block, we find $D_t\simeq v/2$ for procollagen (see S.I.).

The analysis of \fig{fig-exp}c-d provides compelling new evidence that the two cargo molecules studied undergo retrograde transport during their journey through the Golgi apparatus. Indeed, our formalism enables us to determine the average number of inter-cisternal exchange steps experienced by a protein. In a stack with $N$ cisternae, it is equal to $k + k'$ times the average time ($N/v$) spent in the Golgi,  or equivalently to $2ND_t/v$. We thus predict an average of $2N\simeq 10$ exchange steps for VSVG, and $5$ to $10$ steps for procollagen. Since $v>v_t$ and using $k=D_t+v_t/2$ and $k'=D_t-v_t/2$, we find that at least a fourth of these transport steps is backward (toward the ER).

\begin{figure*}[b] 
   \centerline{    \includegraphics[width=17cm]{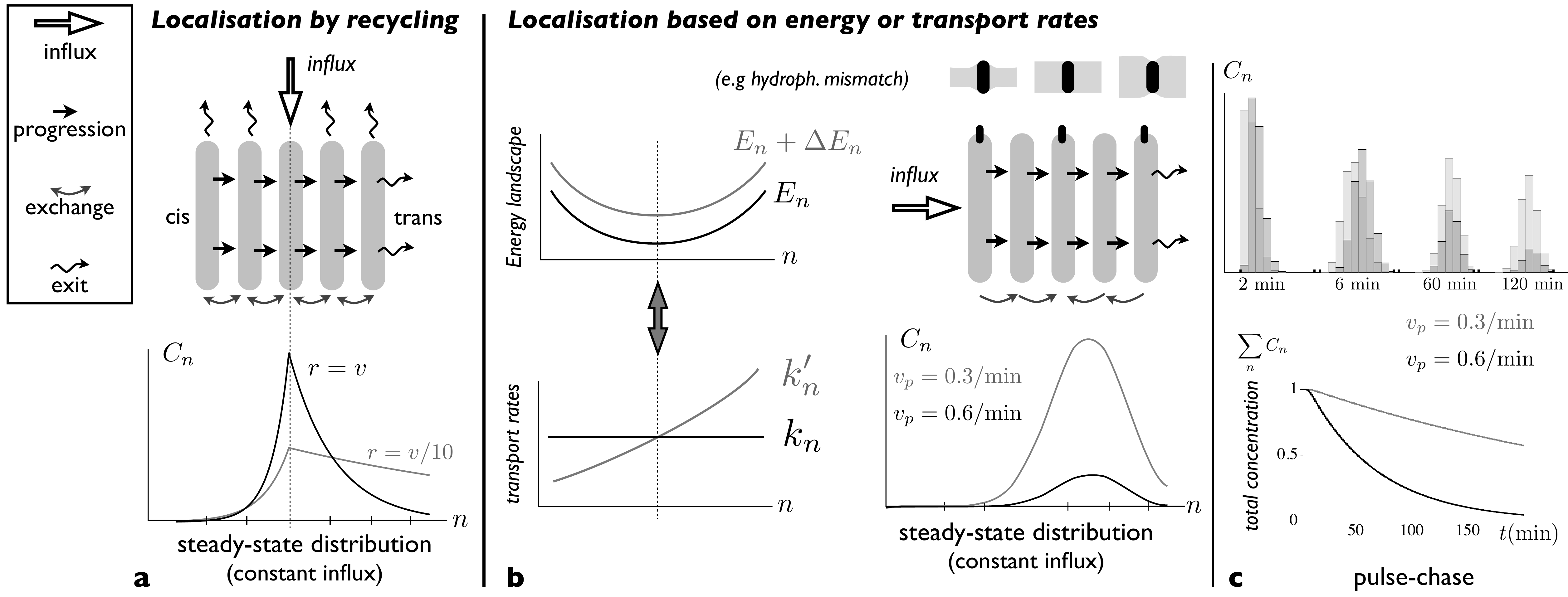} }
   \caption{    \label{fig-resident}
 \small Localization of Golgi resident proteins {\bf a.} Fast recycling of proteins imported at specific Golgi location leads to a peaked protein distribution around the import location. The steady-state distribution profile is shown for parameter values corresponding to VSVG (not a resident protein, $v=D_t=0.3$, $r=v/10$ - grey curve) and for a recycling 10-times faster  (black curve). {\bf b.} Local variation of the transport rates $k$ and $k'$ can be converted into energy landscapes $E,\ \Delta E$ and related to physical mechanisms, such as hydrophobic mismatch. The example shows a quadratic landscape $E_n=K/2(n-n_0)^2$ and the corresponding rates. The steady-state distribution shows a peak where the net velocity $v=k-k'+v_p$ vanishes. {\bf c.} Pulse-chase experiment on resident proteins in a quadratic energy landscape, showing the evolution of a protein distribution initially localised at the {\em cis} face at $t=0$, and the variation of the total protein content with time. Variation of the progression velocity strongly influences the protein distribution and lifetime in the Golgi. Larger $v_p$ (black curve) displaces the peaks toward the {\em trans} face and promotes protein exit.}
\end{figure*}

\subsection{Protein retention inside the Golgi}
Thus far, in our analysis of the transport of transiting proteins, it sufficed to take the transport rates  between Golgi cisternae to be constant and independent of the cisternal index $n$.
We now apply our formalism to resident Golgi proteins (e.g. glycosylation enzymes), that define the identity and function of specific cisternae and thus
must remain in particular locations along the stack. 
Scenarios for protein retention in the Golgi  usually involve either fast recycling of proteins by transport vesicles and/or localization by interaction with the surrounding membrane environment.
A popular mechanism for the latter is the hydrophobic mismatch \cite{webb:1998,fullekrug:1998}, in which proteins are sorted by the span of their transmembrane domains compared to bilayer thickness.
 
In our framework (Eq.\eqref{kineq}), localization by recycling corresponds to an influx of protein targeted to a particular cisterna $n_0$ combined with protein exit at every cisterna ($J_{\rm in}^n=J_{\rm in} \delta(n-n_0)$ and $r>0$). The stationary protein distribution along the stack is the stationary solution to Eq.\eqref{kineq} (see the S.I. for details):
\be
A_\pm(n)=\frac{J_{\rm in}e^{\lambda_\pm(n-n_0)}}{\sqrt{v^2+4rD_t}}\quad,\quad \lambda_\pm=\frac{v}{2 D_t} \pm \sqrt{\frac{v^2}{4 D_t^2}+\frac{r}{D_t}}
\label{lambda}
\ee
The protein distribution is peaked at $n_0$  and is asymmetric, $A_-$ corresponding to $n>n_0$ and $A_+$ to $n<n_0$. It is spread over $1/\lambda_-$  (respectively $1/\lambda_+$) cisternae toward the {\em trans} (respectively {\em cis}) Golgi face, and is broader toward the {\em trans}      face due to protein convection. Accurate protein localization requires $r \ge v$, as illustrated in \fig{fig-resident}a, a much faster rate than the one we measured for the transiting protein VSVG ($r\simeq v/10$). The stationary distribution (Eq.\eqref{lambda} and \fig{fig-resident}a) requires that the influx  is balanced by the outflux $J_{\rm in}=r\int_1^N dn A(n)$, but is not sensitive to details of the recycling pathway. Whether proteins leaving the Golgi are  recycled to cisterna $n_0$ directly or via a more complex pathway ({\em e.g.}  involving the ER or lysosomes) does not modified the steady state profile.

 The effect of the biochemical environment on protein retention corresponds to a variation of the transport coefficients $v_t$ and $D_t$ along the stack. Generically, protein movement in the Golgi can be written as a diffusion in an effective energy landscape $E(n)$ characterizing the protein's energy in the different cisternae, supplemented by an activation energy $\Delta E(n)$ associated to transport intermediates.  In the S.I., we show that :
\begin{eqnarray}
D_t = \frac{k_n}{2} \left( 1+e^{\partial_n E} \right) \quad {\rm and} \quad v_t = k_n \left( 1-e^{\partial_n E} \right) \nonumber \\
{\rm with } \quad k_n = k_0 e^{-\Delta E(n) }\quad {\rm and}\quad k'_{n+1}=k_n e^{\partial_n E(n)}
\label{landscape_rates}
\end{eqnarray}

A landscape that promotes localization near a particular cisterna $n_0$ can locally be written as a quadratic potential: $E(n)=\frac{1}{2}K (n-n_0)^2$, where $K$ is the coupling strength. About half the proteins  moving through such a landscape would be localised at or near the minimum $n_0$ with a spread  $\Delta n=1/\sqrt{K}$ cisternae. Adding a bulk flow ({\em e.g} due to cisternal progression) with velocity $v_p$ displaces the energy minimum from $n_0$ by an amount $\delta n \approx v_p /( K D_t)$ (see \fig{fig-resident}b and S.I.). Thus a large coupling strength $K\gtrsim v_p/D_t$ is required for this localization to be both precise and robust.  

The  landscape approach allows us to test the relevance of the hydrophobic mismatch mechanism, for which the energy $E(n)$ can be computed. The membrane thickness of organelles is known to continuously increase along the secretory pathway from about $3.7$\,nm  in the ER to $4.2$\,nm  at the plasma membrane \cite{mitra:2004} and proteins could be confined to membranes that best match the length of their hydrophobic domains. The energy of hydrophobic mismatch increases quadratically with the thickness mismatch, and leads to a quadratic energy landscape with a coupling strength $K\sim K_s\delta h^2\simeq 0.25\kT$  \cite{phillips:2009} ($\delta h\simeq0.1$\,nm is the mismatch between adjacent cisternae and $K_s\simeq0.1\unit{J/m^2}$ is the bilayer stretching modulus). 
 Hydrophobic mismatch can thus in principle localise proteins against thermal fluctuations with an accuracy of about $\Delta n\sim1/\sqrt{K}=2$ cisternae, and protein localization is indeed known to be affected by the length of its transmembrane domain \cite{machamer1993targeting,rayner:1997}. It is however not robust against variation of the anterograde flux since $K<v/D_t\simeq 1$, consistent with the observation that the transmembrane domain length was not the sole factor affecting protein localization in the Golgi  \cite{machamer1993targeting}.

The two mechanisms above (localization by recycling and by an energy landscape) were used to analyse the distribution of the resident enzyme man I in {\it Arabidopsis} Golgi stacks, see Fig.S2 in the  S.I. This enzyme is localised to cisternae 3 and 4 of the stack with a  $90\%$ accuracy \cite{donohoe:2013}. Such strong confinement requires either fast recycling ($r\simeq 2.6 v$) or a deep energy well ($K\simeq 2.2$). Fast recycling would suggest that Man I is recycled  from all cisternae directly to cisternae 3 and 4, without necessarily leaving the Golgi complex. The large value of $K$ is inconsistent with retention solely based on hydrophobic mismatch ($K\simeq 0.25$), but could stem from an asymmetry in $k_n$ and $k_n'$.

\section{Discussion}

\subsection{Cisternal progression or vesicular transport} 

Our framework produces two strong predictions; (1) the level of inter-cisternal exchange (although not its directionality) can be directly quantified by measuring the coarse-grained  diffusion coefficient $D_t$, and (2) measuring a convection velocity $v>2D_t$ would {\it necessarily} imply some level of cisternal progression. We stress that cisternal progression cannot be disproved in case $v< 2D_t$, since this could correspond to progression combined with retrograde vesicular transport. Our analysis of the data clearly reveals the existence of some degree of diffusion, including significant backward transport steps, {\em both} for the small membrane protein VSVG {\em and} for the large protein complex procollagen (\fig{fig-exp}). Furthermore we find that $v\simeq D_t$ for both species. This implies that (1) inter-cisternal exchange is confirmed in both cases, and (2) cisternal progression cannot be proved nor disproved by the existing transport data. We emphasize that this follows from a strict application  of general transport principles,
and reflects the inadequacy of the existing experimental data to be more discriminating. Detailed microscopic models used to interpret coarse-grained experimental transport data \cite{patterson:2008}  should be viewed with caution, confirming both the utility and necessity of our coarse-grained approach.

 \begin{figure}[b]
\centerline{\includegraphics[width=6.5cm]{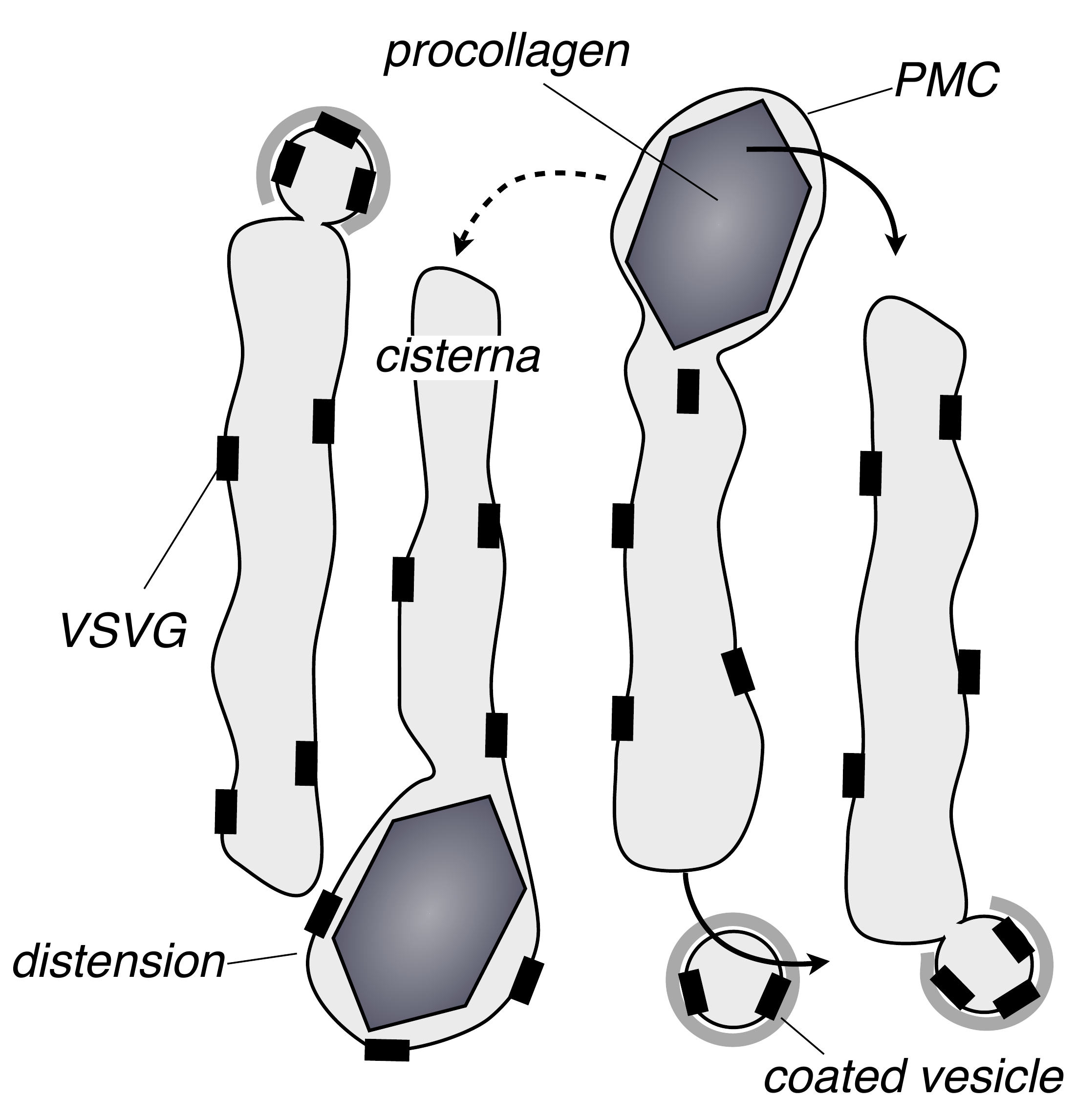}}
\caption{\label{mechanism} { \small Given the evidence for inter-cisternal exchange during the transit of {\em both} small membrane proteins (VSVG) and large protein complexes (procollagen) through the mammalian Golgi apparatus,
 we propose that Pleomorphic Membrane Carriers (PMCs) could be involved in inter-cisternal exchange, possibly triggered by the formation of a protein coat or initiated by distensions in procollagen-containing cisternae. The sketch presents a snapshot of a dynamic  process, showing a PMC being exchanged between two cisternae:  according to our analysis, a procollagen complex is exchanged an average of $6$ times between cisternae during its journey through the Golgi ($\sim 20\unit{min.}$). Note that this exchange process may not lead to a net progression along the stack, and does not invalidate cisterna progression as the main cause for anterograde protein transport.
}}
\end{figure}

One of the main arguments used to support  cisternal progression was the fact that the observed progress of large procollagen aggregates through the Golgi stack was at odds with packaging and transport in  conventional small protein-coated transport vesicles~\cite{bonfanti:1998}. 
However, analysis based on our general framework shows that procollagen is {\em exchanged between cisternae  despite its size}.  This  implies that transport is at least partly mediated by large ``pleiomorphic membrane carriers'' (PMCs) as sketched in  \fig{mechanism}.  PMCs containing procollagen aggregates could take the form of  ``megavesicles'' such as those involved in the transport of large (engineered) protein complexes \cite{volchuk:2000}, or of large tubulo-vesicular connections such as those connecting the Golgi to surrounding organelles \cite{luini2005large}. Such large transport intermediates have not yet been seen experimentally. However, it was recently observed that 
large  protein complexes do progress through the Golgi stack if they are soluble but do not if they are membrane-bound and staple  the two faces of Golgi cisternae  together \cite{lavieu:2013}. This also suggests that large cargo can be exchanged between cisternae. A mechanism based on lateral segregation in the cisternal membranes caused by a Rab cascade, the ``cisternal progenitor model'' \cite{pfeffer:2010b}, has recently linked the formation of large intra-Golgi transport carriers to the maturation of membrane components. The present work is, to our knowledge, the first quantitative evidence of their involvement in intra-Golgi transport directly based on  transport data. Note that inter-cisternal exchange  could be quite fast, so a given procollagen aggregate could only spend a very short fraction of its transit time outside cisternae. If an exchange step takes $\sim1 \unit{sec.}$ and there are $10$ such steps for a transit time of $\sim 15\unit{min.}$,  an aggregate spends about $99\%$ of its time inside cisternae. This suggests that the formation of megavesicles could be a rare event.

Finally, it is intriguing diffusion and convection are found to occur at similar rates ($v\simeq D_t$) for both cargoes. This could indicate that these two processes share the same underlying mechanisms and/or that one process is coupled to the other, as suggested by the cisternal progenitor model \cite{pfeffer:2010b}. More insight could be gained by comparing values for $v$ and $D_t$ in different organisms. Scale-forming algae such as {\it S. Dubia} have a regularly stacked Golgi  of 15-20 cisternae.  Proteoglycan scales readily identifiable by EM and too large to fit in conventional transport vesicles, transit through the stack without ever being seen outside cisternae \cite{donohoe:2013}. The absence of scale-containing megavesicles would imply that  these scales undergo pure convection in the Golgi. A quantitative incoming wave experiment, yet unavailable to our knowledge, should produce data along the dashed lines of  \fig{fig-exp}d, corresponding to the absence of diffusion.

\subsection{Experimental proposal}
We have shown that all available data on the transport of two very different types of cargo through the Golgi (the small membrane protein VSVG and large the collagen complex procollagen) are reproduced by a model of intra-Golgi transport involving constant anterograde and retrograde transport rates, corresponding to a net constant velocity $v$ and constant effective diffusion coefficient $D_t$. Models involving more than these two or equivalent parameters for intra-Golgi transport are not falsifiable by current transport experiments and should be treated with caution. 

Our analysis shows that diffusion, a signature of inter-cisternal exchange, contributes to the transport of {\it both} types of cargo. This is rather surprising for the large protein complex procollagen and should therefore be confirmed by additional transport data with high statistical significance, and using a fast ($< 10$ min) transport block protocol.  We advocate the use of high resolution microscopy  instead of low resolution optical assays (FRAP), since the latter are dominated by the boundary conditions (\fig{fig-exp}a,b) and do not give sufficient insight into the intra-Golgi dynamics.

Direct evidence for cisternal progression may be obtained only if  $v>2D_t$, however our analysis of the transport data showed $v\simeq D_t$ for both types of cargo. More information on the nature of protein transport could be gained by studying  correlation in the transport dynamics of different protein species. A promising technique is the newly developed RUSH method \cite{boncompain:2012}, which allows one to precisely control the release of proteins from the ER into the Golgi, following which their progression and export can be monitored by optical or electron microscopy.

More insight on the interplay between progression and exchange could be gained by comparing the dynamics of transiting and resident Golgi proteins. Monitoring the distribution and dynamics of resident proteins under conditions that affect the  transport of transiting proteins could be a promising strategy, as the localization of resident proteins is affected by cisternal progression. One first needs to identify the mechanism by which particular resident proteins are localised,  fast recycling (\fig{fig-resident}a) and localised retrograde transport related to an energy landscape  (\fig{fig-resident}b) being the two generic ones.
The distribution of resident proteins within the stack should then be determined, by high resolution microscopy, under conditions affecting the transit time of proteins putatively transported by cisternal progression, such as drugs targeting the cytoskeleton. According to our prediction this distribution should correlate with the transport rate of transiting proteins in one of two ways if transit is due to cisternal progression. The distribution of proteins localised by fast recycling should broaden, while the distribution of proteins localised by retrograde transport should be displaced (and not broaden) toward the {\em trans} Golgi face, under conditions that decrease the Golgi transit time (see \fig{fig-resident}). Finding such correlations would bring support to the cisternal progression mechanism.

We close by recalling that transport solely based on cisternal progression cannot be reconciled with existing transport data \cite{patterson:2008}. Exchange mediated by large membrane structures (the PMCs) seems the most reasonable compromise, and can be linked to biochemical maturation by the cisternal progenitor model \cite{pfeffer:2010b}. In fact, the distinction between cisternal progression and inter-cisternal exchange becomes less clear if transport involves large PMCs (\fig{mechanism}) of size possibly close to the cisterna size, that undergo frequent scission and fusion. A more crucial question is rather whether there exist a bulk  anterograde flow of material in the Golgi, or whether transport is mainly protein-specific. Dynamical correlation between different transiting proteins could  inform us on the extent to which they use the same carrier. It would in particular be very interesting if the transport of VSVG, for instance, was increased by the presence of procollagen. That would suggest that procollagen can create its PMCs and that VSVG can be exchanged between cisternae by riding along these structures.

\begin{acknowledgments}
This work was supported by the foundation Pierre-Gilles de Gennes pour la recherche. We acknowledge stimulating discussions with Bruno Goud, Vivek Malhotra, Satyajit Mayor and Franck Perez.
\end{acknowledgments}





\end{article}
\end{document}